# Bounded rationality in a dynamic alternate game

*Eduardo Espinosa-Avila*
eespinosa@uxmcc2.iimas.unam.mx
National University of Mexico (UNAM)

*Francisco Hernández-Quiroz*
fhq@ciencias.unam.mx
National University of Mexico (UNAM)

## ABSTRACT

From the standpoint of game theory, dominoes is a game that has not received much attention (specially the variety known as *draw*). It is usually thought that this game is already solved, given general results in game theory. However, the determination of equilibria is not feasible for the general case because of the well known problem of node explosion in the tree expressing the game. We propose a new model based in limited forecast as a kind of bounded rationality for dynamic alternate games.

## 1. INTRODUCTION

There are a lot of possible games to play with dominoes tiles. The variety we analize here is known as *Draw*, which is part of a bigger group called *blocking games*. In this kind of games, tiles are initially randomly distributed among all players and the one with the biggest *double* (a tile with the same number in both sides) draws it on the table, starting the *game train*. Afterwards, players draw tiles alternately matching a *free-end*. The game ends when one player draws all of her tiles or the game is *blocked*, which happens when none of the players can draw a matching tile. At the end of the game the sum of points in the tiles of the other players are the winner's profit.

Even though there are several theoretical results proposing a solution to similar games, we think this is an interesting problem because in practice it is not computationally possible to apply these results to the general (i.e. with an arbitrary number of points). Also the node explosion problem makes necessary to find alternative techniques to compute the game equilibria and to determine the best strategy in order to get the best profit in the match.

We define a model of the game considering limitations on forecast as referred in [1, 4, 10], but in our approach limitations are not fixed anf instead they are a function of some bounded optimization parameter [5].

## 2. PREVIOUS MODEL

Philippe Jehiel [1] presents a model of limited horizon forecast applied to repeated alternate games. The key features of this class of games are that there are two players moving sequentially in discrete time steps. In each period $t$, the current payoff for player $i$ depends on her own action chosen in that time and the action made by the opponent in the last period. The action spaces are finite and remain the same throughout the match.

It is assumed that each player has a *limited ability to forecast* the future. Player $i$ is characterized by the lenght of her forecast $n_i$ (a constant). At period $t$, player $i$ formulates predictions for the forthcoming $n_i$ moves after her own move. Therefore, she must make her choice of the current action on the basis of her limited forecast *only*. This is because:

1. Player $i$ cannot build her criterion on what will come after $n_i$ periods, since she cannot make predictions about (she has no idea of) it, and,

2. Given the stationarity of the game, the average payoff over the lenght of foresight may be perceived as a good approximation of the true objective function.

Jehiel also defines a solution concept called $(n_1, n_2)-solution$ which requires two preliminary notions:

1. A strategy for player $i$ is *justified* by a sequence of forecasts if the strategy provides actions that maximize the average payoff obtained over the lenght of her forecast, and,

2. A sequence of forecasts for player $i$ is *consistent* with a strategy profile if the forecast coincide with the truncation to the first $n_i$ actions of the respective actions of the respective continuation paths induced by the strategy profile.

Hence a $(n_1, n_2) - solution$ is defined as a strategy profile that can be justified by consistent sequences for players 1 and 2. In other words, in a $(n_1, n_2) - solution$:

1. Current actions are chosen to maximize the average payoff over the lenght of her foresight, and,

2. At any period $t$ where player $i$ must move, her forecasts for the forthcoming $n_i$ actions as a function of her current action are correct.

It should be mentioned that the predictions for the forthcoming $n_i$ actions include her own actions and that the equilibrium forecasts about all these actions are assumed to be correct whatever her current action and not only on the equilibrum path.

Another important issue mentioned by Jehiel is the fact that there is no improvement by incrementing the forecast for player, since the game is cyclic.

## 3. OUR MODEL

In the case of dynamic alternate games, like dominoes, there are substantial differences:

- ⋄ While the movements are alternated, the game is dynamic, meaning that the action space is updated after each move and the search space is reduced.

- ⋄ According to the classification presented in [2], dominoes is a convergent, imperfect information and sudden death game. Given this, the length of the forecast is not constant and forecasting can be better based upon a function of the computing capability of the agent[1].

- ⋄ Since in games with similar nature to dominoes subgame perfect equilibrium can be applied, a reasonable[2] approach to the solution concept could be *subgame perfect equilibrium with limited forecast*, wich would have to compute at each period a new equilibrium according to some desired benefit (payoff function).

- ⋄ A problem in applying the bounded forecast to the game of chess for instance, is the difficulty of determining a *reasonable function* to estimate the payoff at the end of the horizon bounded by the forecast. In the case of dominoes we can use heuristics or *guidelines* known in the folklore of the game to determine this payoff function [7].

In order to develop the model, it is important to have a clear notion of awareness. Therefore, the first step is to answer the questions raised in [3] for dominoes:

- ⋄ Awareness of what? The player must be aware of the actions made up to the current period of the game, as well as the tiles she holds. In addition, she must be aware of the actions she might take in her turn.

- ⋄ What is the environment? The environment consists of the current state of the game: how many tiles each player holds and the game train. A query to the environment consists of trying to reconstruct the history of the match using the current turn and following the match train.

- ⋄ What is the enumeration process? The acquisition of the set of possible actions is made by touring the decision tree of the match as the match evolves. This path can return a set or a particular state. However, building the entire tree requires exponential space.

- ⋄ What is the decision making process? Once the enumeration returns a state or a set, she can select the best possible action from among its outgoing edges by following the subgame perfect equilibrium.

Now we define a model of limited horizon forecast as a function of the computing capability of the agent applied to *dynamic alternate games*. The key feature of this class of games is that there are two players moving sequentially in discrete time steps. In each period $t$, the current expected payoff for player $i$ depends on her own action chosen in that time and the action made by the opponent in the previous period. The action spaces are finite and the search space is reduced as the match evolves.

---

[1]*Computing capability* in this context indicates the ability to generate and visit a certain number of nodes in the future.
[2]*Reasonable* in this context is used as a synonym for common sense.

The latter ensures that in the final steps of the match, the number of nodes is very small and can be determined in reasonable time. In other words, the number of leaves is very small compared to the number of branches at the beginning of the match.

We assume that each player has a *limited ability to forecast* the future. Player $i$ is characterized by her *ability to generate and visit states* in the future $c_i$. At period $t$, player $i$ formulates predictions for the forthcoming $n_i = f(c_i)$ moves after her own move. Therefore, she must make her choice of the current action on the basis of her limited forecast *only*. This is because:

1. Player $i$ cannot build her criterion on what will come after $n_i$ periods, since she cannot make predictions about (she is not aware of) it, and,

2. The subgame perfect equilibrium payoff over the lenght of foresight may be perceived as a good approximation of the true objective function.

## 3.1 Dynamic alternate games

We consider two players indexed by $i = 1, 2$; player $i$ takes actions $a_i$ from a finite action space $A_i$. Players take actions in discrete time and the horizon is finite. Time periods are indexed by $t = 1, 2, 3, ...$. At time $t$ player $i$'s period payoff is a function of the current actions $a_i^t$ of the two players $i = 1, 2$.

Players take actions sequentially and player 1 moves first. At each odd period ($t = 1, 3, 5, ...$), player 1 choses an action from her set. Similarly, player 2 choses her actions at each even period ($t = 2, 4, 6, ...$). In both cases, the action taken modifies the immediate next action of the opponent and reduces the search spaces of both players. We call games like this *dynamic alternate games*.

A stream of action profiles $\{q_i^t\}_{t=1}^{n_{max}} = \{q_1^{2k-1}, q_2^{2k}\}_{t=1}^{n_{max}}$, where $q_1^{2k-1} \in A_1$ and $q_2^{2k} \in A_2$ is known as a path and is denoted by $Q$. Since players move each two periods, an action taken at period $t$ is combined with the action taken by the opponent in the last period $t-1$ to modify the structure of the game tree (they prune it) and therefore, the payoff of player $i$ induced by path $Q$.

## Notation

1. Let $R_n$ be an arbitrary $n$-length stream of actions of alternate actions; $\phi_i(R_n)$ denotes a function that, given the current state for player $i$, returns the expected payoff to player $i$ induced by $R_n$. This function considers both the rules of the game and/or *guidelines* known from the game in question.

2. $[Q]_n$ denotes the truncation of path $Q = \{q_i^t\}_{t=1}^{n_{max}}$, where $n \leq n_{max}$, to the first $n$ actions.

3. $[q]^N$ denotes the truncation of path $q$ to the last $N$ actions.

4. $(q, q')$ denotes the concatenation of $q = \{q_i^t\}_{t=v}^{s}$ with $q' = \{q_i^t\}_{t=s+1}^{w}$: $(q, q') = \{q_i^t\}_{t=v}^{w}$.

## 3.2 The solution concept

Similar to Jehiel [1], we assume that players have a *limited ability to forecast* the future and *bounded recall*; however, unlike his proposal, forecasting ability in our model is not fixed, but a function of the ability of the agent to generate and visit future states in the game tree. The idea of having units of brain power to study the future and partly to the analysis of the past is maintained, but in the case of the units dedicated to the future they are intended to dynamically compute the next possible branches. Therefore, Player $i$ has a two-dimensional ability, on the one hand $N_i$ represents her memory capacity and, on the other side $n_i = f(c_i)$ is the number of steps that player $i$ is able to foresight, as a function of her ability to generate and visit. At each period where player $i$ must take an action, she determines new forecasts about the future. Her forecasts are limited to the next $n_i$ steps. Additionally, as she has bounded memory, her forecasts about the future may only depend on the last $N_i$ periods and her current action.

*Notation and auxiliary definitions*

Let $\mathcal{H}(N_i)$ be the set of histories of alternate actions of length $N_i$, in which last action is an element of $A_j$ $(j \neq i)$ and $h$ an arbitrary element of $\mathcal{H}(N_i)$.

1. An $n_i$-length prediction, where $n_i = f(c_i)$, for player $i$ is a stream of alternate actions of length $n_i$, starting with an action in $A_j$ $(j \neq i)$. The set of $n_i$-length predictions (shorter in the last steps of the match) is denoted by $P_{n_i}$ (a subtree).

2. An $n_i$-length forecast for player $i$ at a period $t$ where she must move is denoted by $f_i^t$. It maps, for every history of length $N_i$, $h \in \mathcal{H}(N_i)$, the set of actions $A_i$ available for the set of predictions $P_{n_i}$. Formally, $f_i^t = \left\{ f_i^t(\cdot|h) \right\}_h$, where, $\forall h \in \mathcal{H}(N_i), f_i^t(\cdot|h) : A_i \to P_{n_i} : f_i^t(a_i|h)$ is the prediction about the forthcoming $n_i$ actions made by player $i$ at period $t$ if she currently choses $a_i$ given the last $N_i$ actions $h \in \mathcal{H}(N_i)$.

3. $f_i = \left\{ f_i^t \right\}_t$ denotes an arbitrary sequence of forecasts $f_i^t$ for every period $t$ where player $i$ must move. The set of $f_i$ is denoted by $\mathcal{F}_i$. A pair $(f_1, f_2) \in \mathcal{F}_1 \times \mathcal{F}_2$ is denoted by $f$ and $\mathcal{F}$ denotes the set of $f$.

4. A pure strategy for player $i$ is denoted by $\sigma_i$. It is a sequence of functions $\sigma_i^t$, one for each period $t$ where player $i$ must take an action. The function at period $t$, $\sigma_i^t$, is the behavior strategy for player $i$ at that period. It determines player $i$'s action at period $t$ as a function of the last $N_i$ actions. A strategy profile $(\sigma_1, \sigma_2)$ is denoted by $\sigma$ and the set of strategy profiles $\Sigma_1 \times \Sigma_2$ is denoted by $\Sigma$.

Any strategy profile $\sigma \in \Sigma$ generates a path $Q(\sigma) = \left\{ q_i^t(\sigma) \right\}_t$, $i = 1$ if $t$ is odd and $i = 2$ if $t$ is even. Let $\mathcal{H}^t$ be the set of histories of alternate actions of length $t$ and let $h^*$ be an arbitrary history of length $t-1$, i. e., $h^* \in \mathcal{H}^{t-1}$. The strategy profile and the induced path by $\sigma$ on the subgame $h^*$ are denoted by $\sigma|_h$ and $Q(\sigma|_h)$ respectively. Given $h^* \in \mathcal{H}^{t-1}$ and an action $a_i \in A_i$ at period $t$, the continuation path induced by $\sigma$ after $(h^*, a_i)$ is thus $Q(\sigma|_{h^*a_i})$. The set of continuation paths at period $t$, referred as the continuation set, is denoted by $Q^t(\sigma) = \{(a_i, Q(\sigma|_{h^*a_i}))\}_{h^*a_i}$. The sequence of continuation sets $Q^t(\sigma), t = 1, 2, \ldots$ is denoted by $\hat{Q}(\sigma) = \left\{ Q^t(\sigma) \right\}_t$.

The key idea in this construction is that the strategies of player $i$, (i. e. her choices of actions), are based on her forecast, limited by her computing capability. Hence, to define the solution concept it is necessary to (1) specify a criterion based on forecast as a function of computing capability and (2) show how equilibrium forecasts are related to equilibrium strategies.

The criterion for calculating the payoff is to determine the largest profit among all the *branches* in the subtree she can see by applying subgame perfect equilibrium. Such a criterion is natural, given the features of this class of games. In the early stages (where the player cannot see the horizon of the match) this criterion will indicate what action will give her the best profit even if this profit may not be the best in the whole game tree. In the final stages (where player can see all subtrees from the current node) the agent can determine the *actual equilibrium* and take the *best actual action* according to the subgame perfect equilibrium. In other words, player assumes she can see the whole tree and based on this assumption, she can calculate a subgame perfect equilibrium at every turn. Formally:

**Definition 1.** A strategy $\sigma_i \in \Sigma_i$ is *justified* by a sequence of forecasts $f_i = \left\{ f_i^t \right\} \in \mathcal{F}_i$ if at each stage player $i$ chooses the action that delivers the largest profit by applying subgame perfect equilibrium to the game tree produced by her forecast limited by her computing capability.

We next assume that player $i$'s equilibrium forecasts are related to equilibrium strategies by a *consistency relationship*, defined as follows. Given a history $h^*$ of length $t - 1$, and any action $a_i$ at the current period $t$, $Q(\sigma|_{h^*a_i})$ is the continuation path induced by $\sigma$ after $(h^*a_i)$. At period $t$, the $N_i$ last actions are $h = [h^*]^{N_i}$. Consistency requires for every $(h^*, a_i)$, the forecast $f_i^t(a_i|h)$ coincides with the truncation to the first $n_i = f(c_i)$ actions of the continuation path induced by $\sigma$, $[Q(\sigma|_{h^*a_i})]_{n_i}$. In other words, consistency means that forecasts are correct on and off the equilibrium path. Formaly:

**Definition 2.** $f_i = \left\{ f_i^t \right\} \in \mathcal{F}_i$ is *consistent* with $\sigma \in \Sigma$ if for every period $t$ where player $i$ must move: $\forall a_i \in A_i$, $\forall h^* \in \mathcal{H}^{t-1}$, $f_i^t(a_i|h) = [Q(\sigma|_{h^*a_i})]_{n_i}$ with $h = [h^*]^{N_i}$.

Now define the solution concept. A $(c_1, c_2) - solution$ is a strategy profile that can be *justified* by *consistent* forecasts for players 1 and 2, i. e., a strategy profile that is associated with sequences of forecasts such that (1) players choose their actions in order to maximize the payoff received by applying the subgame perfect equilibrium over the length of her current forecast and (2) player $i$'s forecasts for the forthcoming $n_i = f(c_i)$ actions after her own move are correct on and off the equilibrium path. Formally:

**Definition 3 The solution concept.** A strategy profile $\sigma = (\sigma_1, \sigma_2) \in \Sigma$ is a *subgame perfect equilibrium with limited forecast* ($(c_1, c_2) - solution$) if and only if there exist sequences of forecasts $f = (f_1, f_2) \in \mathcal{F}$ such that for $i = 1, 2$.

1. $\sigma_i$ is *justified* for $f_i$ and

2. $f_i$ is *consistent* with $\sigma$.

Similarly to Jehiel, we do not provide justification for why forecasts should be correct in equilibrium, but in [8] Jehiel discusses a learning process based on limited predictions such that players eventually learn to have correct forecasts.

Hence, players eventually behave as in $(c_1, c_2) - solution$.

## 3.3 Construction

Given a forecast $f_i^T$ of player $i$ at period $T$, if $f_i^T$ is associated with a $(c_1, c_2) - solution$, is it possible to derive $f_i^{T-1}$ backward on the sole basis of $f_i^T$? It is not possible, because player $i$ generates subtrees at each period $T$ where she must move, hence the opponent cannot know in advance which tree she will generate (since it depends on her computing capability). There are two ways in wich the opponent may foresee all the moves down to the leaves: if she has exponential capability and when she is *sufficiently near* to the horizon of the match; if this is the case, she will be able to determine a subgame perfect equilibrium.

Hence in dynamic alternate games the construction is performed forward, but in each period $T$ where player $i$ moves, she must apply backward induction to calculate the subgame perfect equilibrium over the generated subtree. This construction is similar to the *forward looking procedure* introduced in [9].

As stated above, on each step player $i$ generates a set of predictions $P_{n_i}$ of length (depth) $n_i = f(c_i)$, this set is the subtree computed at the current period. Given $P_{n_i}$ we apply a function that returns a payoff for each outgoing edge from the current node to the *leafs* of each prediction. When player $i$ cannot see the *real* horizon of the game from the current node this function estimates the gains from the rules and/or *guidelines* known from the game in question. Once the player can see the whole subtree starting from the current node, the function returns the payoff for each of the leafs. Therefore the player can obtain a series of *subgame perfect equilibria* wich corresponds with a $(c_1, c_2) - solution$.

Moreover, this function is useful to make the best choice at each time step and to generate a sequence of subtrees $\mathcal{P}_{n_i}$. Each $P_{n_i} \in \mathcal{P}_{n_i}$ meets that $length(P_{n_i}) \leq length(P_{n_{i+1}})$ where $n_i$ is such that the current player cannot see the leafs. Once the player is able to observe the entire subtree from any node, the relationship is reversed $length(P_{n_i}) \geq length(P_{n_{i+1}})$, since the number of nodes in the last steps of the game tree is much lower than in the early stages due to the considerable reduction of the search space as the game progresses.

## 3.4 Properties

A dynamic alternated game always has at least one subgame perfect equilibrium with limited forecast. By applying a Kuhn's Corollary of the Zermelo-von Neumann's Theorem [10], we guarantee the existence of a subgame perfect equilibrium for each subtree that player $i$ may generate, hence we may construct a subgame perfect equilibrium with limited forecast by concatenating the equilibria calculated in each period.

Equilibrium forecasts associated with $(c_1, c_2) - solution$ are history independent, as a decision made in the current period is taken on the sole basis of the last action (taken by the opponent).

## 3.5 When a player is better off with a larger foresight

Given the nature of dynamic games, the search space is reduced at each period and since the game is finite (of sudden death), contrary to the model developed by Jehiel [1], in this class of games player $i$ gets advantage by having larger computing capability and, therefore a larger foresight.

Compared to a completely random player, this model behaves better. At any stage, while a purely random player chooses her actions completely at random within the range of possible options, a player implementing the model presented here can take into account both knowledge and preferences of the player and other players (the guidelines).

On the other hand, this model requires less computational power than that of "rational man". Endowed with polynomial capability, the player's behaviour approaches that of the rational man as the game advances.

## 3.6 Example

To illustrate the usefulness of the model, we now present how to apply it to a small instance of dominoes. As stated above, the problem with games like chess is the difficulty to define a payoff function. However, in games like dominoes we may use basic guides like those shown in [7] to define the expected payoff function $\phi$. Some of them can easily be adapted to the case of individual games.

We consider an instance of players $\{1, 2\}$ with 6 tiles (each player gets 3 at the beginning). To show a concrete example, player 1 gets tiles $\{(0,0), (0,1), (2,2)\}$ and player 2 $\{(0,2), (1,1), (1,2)\}$; additionally, the computing capability for each player is $2n$ with $n$ the number of tiles each player gets at the beginning. To simplify the example we will consider the guide of drawing the tile with higher value. The main goal of the match is to win by drawing all the tiles she possesses first and if she cannot win in the current period she will apply the guide mentioned above. It should be noticed that when nodes generated by an agent do not cover a full level, we consider the she does not possesses any information about the truncated level, as we cannot know what set of nodes she will generate.

The game described above generates the game tree shown in figure 1. We observe that if player 1 draws the tile $(0,0)$ in her first turn, there exists a path that gives her a gain of 3. However, as she has limited forecast she is not aware of this possibility. At the beginning the number of levels player 1 may visit is $n_1 = 1$; therefore she evaluates her profit with the guide of "getting rid of heavier tokens" and she draws tile $(2,2)$.

Figures show in boxes the levels that player $i$ can generate and visit at each turn and in light gray the branches cannot be played. Figure 3 shows that player 2 can foresee $n_2 = 2$ levels down.

Now player 2 has two options: draw tiles $(0,2)$ and $(1,2)$. If she draws tile $(0,2)$, she might win the match; but as she is not aware of it she chooses to draw tile $(1,2)$, the highest.

At this point, figure 3 shows that player 1 can only draw tile $(0,1)$. She can see the complete rest of the game tree $n_1 = 4$ and she is aware she will win, getting a gain of 2.

Figure 4 shows the final stages of the match. Player 2 must draw tile $(0,2)$ but she can leave free-end $[0,0]$ or $[2,2]$ in both cases she loses. She chooses to leave $[2,2]$ in order to *lock* the tile $(0,0)$ of player 1.

This example shows that the length of the forecast grows as the game evolves because the number of future states is reduced at each period. This feature together with the intrinsic finiteness of the game makes feasible to build a subgame equilibrium (and strategies) that *gets closer* to the perfect at each step.

**Figure 1: Full game tree**

**Figure 2: Game tree after the first turn of player 1**

**Figure 3: Game tree after the first turn of player 2**

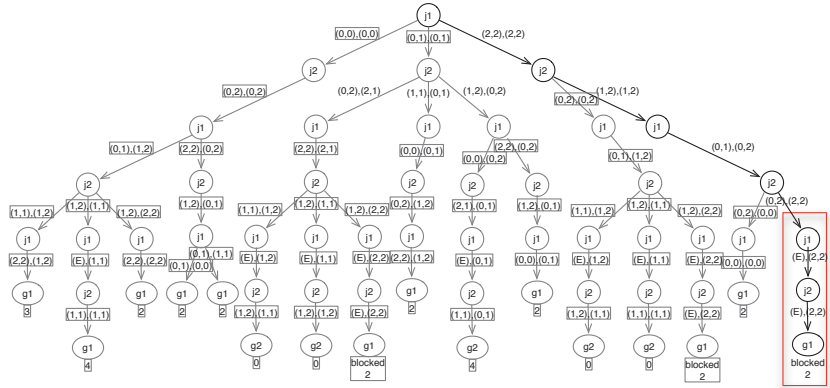

**Figure 4: Game tree in the final stages of the match**

## 4. CONCLUSIONS AND FURTHER WORK

In this paper we have proposed a model of limited foresight as a function of the computing capability of an agent applied to the class of dynamic alternate games. We also define a solution concept called subgame perfect equilibrium with limited forecast which uses the well known subgame perfect equilibrium for each subtree that player builds at each period of the game.

Additionally, we provide a couple of properties for the class of games as well as the solution concept. Finally, we show a concrete example of the model applied to an instance of the game dominoes in order to show its applicability.

We intend to extend the results in this paper in the following directions:

- ◇ Currently we seek to prove that our model performs better than a purely random one.
- ◇ Develop a generic mathematical model applicable to a class of games, not only dominoes.
- ◇ Derive other interesting properties about the solution concept.
- ◇ Present concrete examples on how to use the model, initially by applying it to dominoes adding several guides and then with other dynamic games.
- ◇ Obtain models that combine other characterizations of bounded rationality.

# Appendix

## Kuhn's Corollary of the Theorem of Zermelo-von Neumann

A general n-person game $\Gamma$ with perfect information always has an equilibrium point in pure strategies [10].

## Basic strategies for dominoes

Here are descriptions of basic moves and strategies [7].

1. **Commanding/Leading a Strong Number.** Generally a player should lead a strong number with the objective of playing it later in the game. A strong number is a number that occurs often in a player's hand.

2. **Indicating/Showing the Number of the Double Tile.** A player should command a number that includes any respective double tile in her hand, so that her partner knows her most difficult tiles to play.

3. **Hitting/Blocking a Number Commanded by the Opposition.** When a player blocks a number led by the opposition.

4. **Leaving a Number Open.** When a player avoids drawing a number that he has been trying to play.

5. **Repeating a Number.** A player should repeat a strong number.

6. **Taking Care of the Hand.** When a player avoids being void at a given number.

7. **Avoid Leading an Orphan Number.** Generally, leading an orphan number should be avoided because the player who does so is providing inaccurate information to her partner. An orphan number is a number that occurs only once in a hand.

8. **Protecting Your Partner/Avoiding a Possible Pass.** If a player does not have the relative control of her couple, she should avoid forcing her partner to pass on her next turn.

9. **Indicating/Showing Your Type of Hand.** Each player should try to show the value of her hand (low or high) so that her partner knows the tiles she should try to play.

10. **Stealing the Game.** When a player does not have the relative control of her couple, she should hit a number commanded by her partner with her own strong number.

11. **Playing Aggressively (for Low Hands).** A player should try to play in a manner that high tiles cannot be played if she feels that her couple can win the game or she does not have a high hand with at least one high double hand.
    From this strategy comes the first objective of dominoes: try to win by getting the **highest** amount of points.

12. **Playing Conservatively (for High Hands).** A player should try to play in a manner that high tiles are played if she feels that her couple cannot win the game or she has a high hand with at least one high double hand.
    From this strategy comes the second objective of dominoes: try to lose by giving the **lowest** amount of points.

13. **Playing to Accumulate Points.** The couple has the option to play aggressively (for low hands) if the score is not close to the upper limit of points.

14. **Playing not to Accumulate Points.** The couple should play conservatively (for high hands) if the score is close to the upper limit of points. The definition of a "close" score is subjective and depends on the player's appraisal.